\documentstyle[12pt]{article}

\begin{document}
\pagenumbering{arabic}
\normalbaselines
\vskip -10pt
%********************************************************************
\baselineskip=.175in
%******************************************************************

\noindent
{\bf Anderson} {\it et al.} {\bf reply (to the comment 
of Murphy on 
``Indication, from Pioneer 10/11, Galileo, and Ulysses Data, 
of an Apparent Anomalous, Weak, Long-Range Acceleration").}

%********************************************************************
\baselineskip=.175in
%******************************************************************
\begin{quotation}
We conclude that Murphy's proposal (radiation of the power of the main-bus 
electrical systems  from  the rear of the craft) can not explain the 
anomalous Pioneer acceleration.
\end{quotation}
%********************************************************************
\baselineskip=.33in
%******************************************************************

In his comment \cite{murphy} 
Murphy  proposes that the 
anomalous acceleration seen in the Pioneer 10/11 spacecraft 
\cite{anderson} can be ``explained, at least in part, by non-isotropic 
radiative cooling of the spacecraft."  So, the question is, does 
``at least in part" mean this effect  comes 
near to explaining the anomaly?  We argue it does not  \cite{us}. 

Murphy considers radiation of the power of the main-bus electrical 
systems  from  the rear of the craft.   
For the Pioneers, the aft has a louver system, and 
``the louver system acts to control the heat 
rejection of the radiating platform...A bimetallic spring, thermally
coupled radiatively to the platform, provides the motive force for 
altering the angle of each blade. In a closed position the heat rejection 
of the platform is minimized by virtue of the ``blockage'' of the 
blades while open louvers provide the platform with a nearly 
unobsructed view of space.''    \cite{piodoc}

If these louvers were open, then, Murphy calculates this would produce 
an acceleration $a_0=9.2\times 10^{-8}$ cm s$^{-2}$.  
Murphy uses numbers for thermal
radiation that correspond to the position of the spacecraft near
Jupiter, i.e., 5.5 AU. At that time, the spring temperature 
was about 56 $^{\circ}$ F, meaning the opening angle of the 
louvers was down to  
20$^{\circ}$. This reduces his estimate for the effective  $a_0$ to
$a\equiv\sin(20^{\circ}) a_{0}=3.2\times 10^{-8}$ cm s$^{-2}$.

However, our effect could  only be seen well beyond 
5.5 AU; i.e., further than 10-15 AU.  By 9 AU the actuator spring 
temperature had already reached $\sim$40$^{\circ}$. This means the 
louver doors were   closed  (i.e.,  the louver angle was zero) 
from there on out. 
Thus, from our quoting of the radiation properties above, 
any contribution of the thermal radiation to the Pioneer anomalous
acceleration should be small. (Certainly it would not 
be expected to be higher 
than it was at a 20$^{\circ}$ opening angle \cite{notes}.)   

%***********************************

In 1984 Pioneer 
10 was at about 33 AU and the power was about 105 W.  (Always reduce 
the effect of the total power numbers by 8 W to account 
for the radio-beam power.)
In (1987, 1992, 1996) the craft was at $\sim$(41, 55, 65) AU
and the power was $\sim$(95, 80, 70) W. 
The louvers were inactive.  No decrease in $a_P$ was seen.

%***********************************

We conclude that this proposal can not explain the anomalous Pioneer 
acceleration.  
 
Heat radiation should be a more significant systematic for Ulysses than 
for the Pioneers.  However, in principle this could be separated out since 
accelerations along the lines of sight towards the Earth 
and towards the Sun could 
be differentiated.  This is one of the reasons why a detailed 
calculation of the Ulysses orbit from near Jupiter encounter to Sun 
perihelion was undertaken, using CHASMP.  

This turned out to be a much more difficult calculation than imagined.  
Because of a failed nutation damper,  an inordinate number of
spacecraft maneuvers were required (257).  
Even so, the analysis has now been completed.  
The results are disheartening.  For an unexpected 
reason, any fit is not significant.   The anomaly is dominated by 
(what appear to be)  gas leaks.  
That is, after each maneuver the measured anomaly changes.  
The measured anomalies randomly change sign and magnitude.  The values 
go up to about an order of magnitude larger than $a_P$.
So, although the Ulysses data was useful for range/Doppler checks 
to test models, like Galileo it could not provide a good 
number for $a_P$.  

The gas leaks so far found in the Pioneers are about an order of 
magnitude too small to explain $a_P$.  
Even so, we feel that some systematic or combination of 
systematics (such as heat or gas leaks) 
will most likely explain the anomaly.  However, 
such an explanation has yet to be demonstrated.  

This work was supported
by the Pioneer Project, NASA/Ames Research Center,
and was performed at the Jet Propulsion Laboratory, California Institute 
of Technology, under contract with 
NASA. P.A.L. and A.S.L. acknowledge support by a grant from NASA
through the Ultraviolet, Visible, and Gravitational Astrophysics Program.
M.M.N. acknowledges support by the U.S. DOE.

%********************************************************************
\vskip 20pt
\baselineskip=.175in
%******************************************************************

\noindent John D. Anderson,$^a$ Philip A. Laing,$^b$ Eunice L. Lau,$^a$ 
Anthony S. Liu,$^c$ Michael Martin Nieto,$^{d}$ and Slava G. Turyshev$^a$

%********************************************************************
\vskip 10pt
\baselineskip=.175in
%******************************************************************

\noindent {$^{a}$Jet Propulsion Laboratory, California Institute of 
Technology, Pasadena, CA 91109}    \\
\noindent {$^b$ The Aerospace Corporation, 2350 E. El Segundo Blvd., 
El Segundo, CA 90245-4691} \\
\noindent {$^c$ Astrodynamic Sciences, 2393 Silver Ridge Ave., Los Angeles, 
CA 90039} \\
\noindent {$^d$ Theoretical Division (MS-B285), Los Alamos National 
Laboratory, University of California, Los Alamos, NM 87545}  

%***********************************************
\vskip 10pt
%****************************************
\noindent Received \today 
%**********************************************
\vskip 10pt
%********************************************************************
\baselineskip=.33in
%******************************************************************

%********************************************************************


\begin{thebibliography}{44}

\bibitem{murphy} E. M. Murphy, previous comment and eprint gr-qc/9810015.

\bibitem{anderson} J. D. Anderson, P. A. Laing, E. L. Lau, A. S. Liu,  
M. M. Nieto, and S. G. Turyshev, Phys. Rev. Lett. {\bf 81}, 2858 (1998). 
For further details see eprint gr-qc/9903024.

\bibitem{us}  From the present wording 
of his comment, it appears Murphy does not disagree too strongly with 
this statement.Indeed, in a private communication from JDA to Murphy 
on 6 Oct. 1998, it was pointed out 
that the Pioneers have louvered doors and not fins as 
radiators.  This, by itself, obviated the ``prosaic explanation" of
the original eprint \cite{murphy} by a large factor. 

\bibitem{piodoc} Pioneer Project NASA/ARC document No. PC-202.

\bibitem{notes} Any change of the louver angle should result
in a spin change due to the thermal radiation. This is because  
of the orientation of the lovers around the bus on the spacecraft.
We detect no such change.


\end{thebibliography}
\end{document}